\documentclass{fizik}%
\usepackage{multicol,epsfig}
\usepackage{bbm}
\usepackage{amsmath} 
\vol{27}
\pyear{2003}
\received{August 6, 2003}

\author[SEELIG, PILGRAM,  B\"UTTIKER]{
\textbf{Georg Seelig, Sebastian Pilgram, and Markus B\"uttiker}\\
\textit{D\'epartement de Physique Th\'eorique, 
 Universit\'e de Gen\`eve,}\\
\textit{CH-1211 Gen\`eve 4, Switzerland}\\
}

\title{Decoherence in ballistic mesoscopic interferometers}

\begin{document}
\maketitle

\begin{abstract}
We provide a theoretical explanation for two recent experiments on
decoherence of Aharonov-Bohm oscillations in two- and multi-terminal ballistic rings.
We consider decoherence due to charge fluctuations and emphasize the role of charge 
 exchange  between the system and the reservoir or nearby
 gates. A  time-dependent scattering matrix approach is shown to be 
a convenient tool for the discussion of decoherence in ballistic conductors. 

\end{abstract}

\section{Introduction}
Mesoscopic rings and interferometers are useful tools for the investigation
of phase coherence and decoherence of electrons. Several recent experiments
have been dedicated to the problem of decoherence in mesoscopic Aharonov-Bohm (AB)
geometries. Hansen et al.~\cite{hansen} investigated decoherence in a two-terminal
mesoscopic ring with only a few ($\sim 3-10$) channels open in the ring.
In their experiment, they measured a decoherence rate  linear in temperature.
Kobayashi et al.~\cite{kobayashi} investigated the measurement-configuration dependence
of the decoherence rate in a mesoscopic ring with four terminals where two terminals 
served as current probes and two terminals were voltage probes. They found a strong 
dependence of the decoherence rate on the probe configuration. The theory presented
here can explain the experimental observations of Refs.~\cite{hansen} and \cite{kobayashi}.
More recently, a different type of electronic interferometer based on the use of quantum Hall 
edge states was experimentally realized by Ji et al.~\cite{ji}
and used to investigate the role of coherence in shot noise \cite{marquardt}. 
Decoherence in ring-like ballistic mesoscopic  structures is also of importance 
in the context of the detection of entanglement in electric conductors. Specifically, ring-like
 structures are used for detection of orbital entanglement \cite{samuelsson1,beenakker,gisin,samuelsson2}.
For a recent review of dephasing in mesoscopic physics see Ref.~\cite{lin}.

 In a typical experimental setup, metallic  gates are used to define the ring 
and/or to control the number of transport channels in the arms and contacts of the ring.
 Since an open mesoscopic conductor can easily exchange charge with neighbouring reservoirs
 the ring can temporarily be charged up relative to the nearby gates. Such charge 
fluctuations  give rise to fluctuations of the internal electrostatic potentials.
In two previous publications \cite{seelig1,seelig2} we have discussed decoherence
of AB oscillations in open ballistic mesoscopic rings due to scattering of electrons 
from potential fluctuations. This  approach is  similar in spirit to the 
one proposed in Ref.~\cite{aak} where dephasing in metallic diffusive conductors
was addressed. Here we will again emphasize the close connection between charge 
fluctuations and decoherence  and present a theory which emphasizes the role of
contacts and nearby gates.  
\begin{figure}
\begin{center}
\includegraphics[scale=0.45]{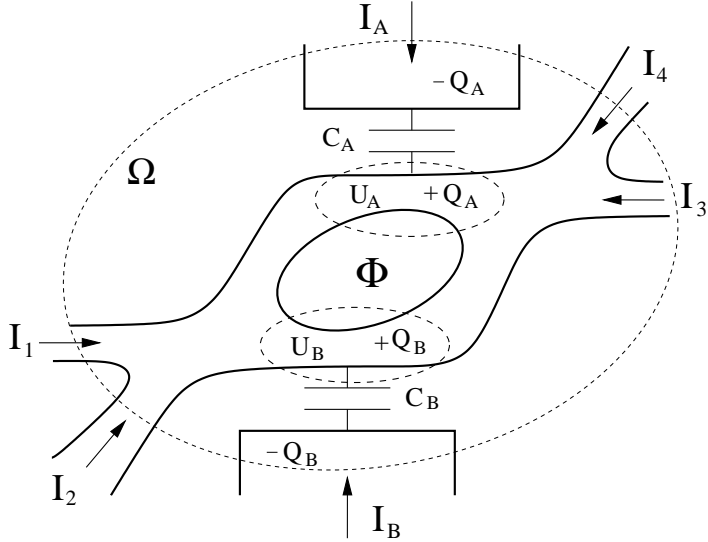}
\end{center}
\caption{Four-terminal Aharonov-Bohm interferometer threaded by a magnetic flux. 
The two arms of the ring are  coupled to metallic  gates via the
capacitances $C_i$  ($i=A,B$). We consider junctions which are 
perfectly transmitting 
and divide the incoming current into the upper and lower 
branches 
of the ring. The total 
charge in a Gauss  sphere $\Omega$ drawn around the system 
of gates and ring is assumed to be zero, implying that current 
in the system is conserved. 
It is assumed that each arm is 
characterized by its internal potential $U_A$ (or  $U_B$)
and its charge  $+Q_A$ (or  $+Q_B$). After Ref.~\cite{seelig1}, copyright (2002) by the American Physical Society.}
\label{mzi_figure}
\end{figure}

\section{An electronic interferometer}
As a first example we will discuss decoherence of Aharonov-Bohm oscillations in a
mesoscopic interferometer with four terminals \cite{seelig1} pierced by a magnetic flux $\Phi$.
 The arms of the ring are coupled to metallic gates. 
A sketch of the system we want to consider is shown in Fig.~\ref{mzi_figure}.
The charge densities $Q_i(x,t)$ ($i=A,B$) 
and therefore the internal potentials $eU_i(x,t)$ fluctuate in time, as charge is exchanged 
between the ring and the reservoirs or between different regions in the ring. Here, the index $A$
 refers to the upper and the index $B$ refers to the lower arm. 
In our theoretical treatment of charge-fluctuation-induced decoherence we make two main 
 assumptions: i) There is no backscattering
 in the intersections where the ring is connected to the external contacts and consequently
 the interferometer does not exhibit closed orbits. This makes the system an electronic
equivalent of the optical Mach-Zehnder interferometer (MZI). ii) Electron-electron interactions
 take place in the arms of the interferometer. They are weak and cause no $2k_F$-backscattering
 or inter-channel scattering. 

In accordance with our first assumption the intersections are described as reflectionless 
beam splitters (see inset in Fig.~\ref{rate_fig})  with a scattering matrix 
\begin{equation}\label{beamsplitter} 
{\bf S}_B=\left(\begin{array}{ll} 
           0&{\bf s}\\ 
           {\bf s}&0 
          \end{array}\right),\,\,\,\, 
{\bf s}=\left(\begin{array}{cc} 
           \sqrt{{\cal T}}{\mathbbm 1}_N&i\sqrt{1-{\cal T}}{\mathbbm 1}_N\\ 
           i\sqrt{1-{\cal T}}{\mathbbm 1}_N&\sqrt{{\cal T}} {\mathbbm 1}_N
          \end{array}\right). 
\end{equation} 
Here, $\sqrt{{\cal T}}$ is the amplitude for going straight through the intersection  
 and $i\sqrt{1-{\cal T}}$ the amplitude for being deflected. It is assumed that 
all contacts and both arms of the interferometer carry $N$ transport
channels and no channel-mixing occurs in the intersections. Furthermore,
transmission through the intersection region is assumed to be instantaneous
or, at least, very fast compared to the traversal times of the leads.

Electron-electron interactions  in the arms of the rings are accounted for 
through the introduction of the time-dependent 
internal potentials $eU_i(x,t)$ ($i=A,B$) which have to be determined self-consistently. 
Electrons passing through the ring scatter from these  potential fluctuations.
In our previous work \cite{seelig1} we treated inelastic transitions with 
the help of a scattering matrix $S(E',E)$ which depends on the energy of incoming and
 outgoing carriers \cite{ap}. In this case, the amplitude $\hat a_{\beta m}(E')$
 of the incoming current in channel $m$  of contact $\beta$ at energy $E'$ and
the amplitude $b_{\alpha n}(E)$  of the outgoing current at contact $\alpha$,
 channel $n$ and energy $E$ are related through $\hat b_{\alpha n}(E)=\sum_{\beta m}\int_{-\infty}^\infty dE'S_{\alpha\beta, nm}(E,E')\hat a_{\beta m}(E')$.

Alternatively, one can also use scattering matrices with two time arguments
instead of two energy arguments. Here, we will follow this alternative approach
 since it allows for a rather convenient formulation or the problem of decoherence
 in mesoscopic rings. The scattering matrix $S(t,t')$ then relates current amplitudes
at different times: \cite{polianski}
\begin{eqnarray}\label{s1}
\hat b_{\alpha n}(t)&=&\sum_{\beta m}\int_{-\infty}^\infty dt'S_{\alpha\beta, nm}(t,t')\hat a_{\beta m}(t'),\\
\hat b^\dagger_{\alpha n}(t)&=&\sum_{\beta m} \int_{-\infty}^\infty dt'S^\ast_{\alpha\beta, nm}(t,t')\hat a^\dagger_{\beta m}(t').\nonumber
\end{eqnarray}
Here, $\left[S^\dagger(t',t)\right]_{\beta\alpha,mn}=S^\ast_{\alpha\beta, nm}(t,t')$.
Note again that the two different formulations of scattering matrix theory
in the presence of inelastic scattering are completely equivalent. The scattering matrices
$S(E',E)$ and $S(t',t)$ are related through Fourier transformation and it is a question of
convenience which approach to use for a given problem. 
Time-dependent scattering matrix theory was applied  to chaotic quantum dots 
in Refs.~\cite{polianski} and \cite{vavilov}. 

Inelastic scattering in the arms of the  MZI is taken into account 
through  the scattering matrix  $S_i(t,t')$ ($i=A,B$) which we will  derive
 in Sec.~\ref{wire}. The scattering matrix
 $S_i(t,t')$ is an $N\times N$ diagonal matrix (no inter-channel scattering).
 The scattering matrix for the full MZI has the dimension $4N\times 4N$ and can 
 be  obtained from the scattering matrices Eq.~(\ref{beamsplitter}) for the beam splitters
 and the scattering matrices for the arms of the ring.  As an example  consider the matrix
 element $S_{31}(t,t')$ which becomes
\begin{equation}\label{s31}
S_{31}(t,t')=i\sqrt{{\cal T}(1-{\cal T})}\left[S_A(t,t'){\rm e}^{i\Phi_A}+S_B(t,t'){\rm e}^{-i\Phi_B}\right].
\end{equation}
Here $\Phi_i$ is a magnetic phase and the positive (negative) sign is for an
electron going through the arm $i$ of the ring in the (counter-) clockwise direction.
Note that $\Phi_A+\Phi_B=2\pi\Phi$ where $\Phi$ is the flux through
 the ring in units of the flux quantum.
The other elements of the total scattering matrix can be 
constructed in a similar way using Eq.~(\ref{beamsplitter}) and the scattering
matrices for the arms.

\subsection{Scattering Matrix for a Wire}\label{wire}
Our next task is to determine the scattering matrix $S(t,t')$ for an electron in
a wire of length $L$ in the presence of a fluctuating time-dependent potential $eU(x,t)$.
Assuming that the problem is separable, we start from the one-dimensional Schr\"odinger equation
\begin{equation}\label{schrodinger}
i\hbar\frac{\partial}{\partial t}\Psi_n(x,t)=\left(-\frac{\hbar^2}{2m}\frac{\partial^2}{\partial x^2}+E_n+eU(x,t)\right)\Psi_n(x,t)
\end{equation}
describing the motion along the wire direction for an electron in the $n$-th transverse channel.
Here, $E_n$ is the cut-off energy for the $n$-th channel, at $E_F<E_n$ the channel is closed. 
 The electrostatic potential $eU(x,t)$ is the same for electrons in different transverse channels.
To find the scattering matrix we are looking for a special solution to Eq.~(\ref{schrodinger})
corresponding to a right-moving particle localized at $x=0$ at time $t=0$.  
 For the wave function we make the ansatz 
\begin{equation}\label{wavepacket}
\Psi_n(x,t)=\int \frac{dk}{2\pi}\psi_n(k;t){\rm e}^{ikx},\,\,\,\, \psi_n(k;t)={\rm e}^{i\phi_n(k;t)/\hbar}.
\end{equation}
If we impose the initial condition $\phi_n(k;t=0)=0$ we get $\Psi_n(x,0)=\delta (x)$.
 The delta function approximates a right-moving wavepacket, extended on the scale set 
by the Fermi wavelength $2\pi/k_F$ but well-localized 
compared to the characteristic extensions of the interferometer. 
Substituting the ansatz Eq.~(\ref{wavepacket}) into the Schr\"odinger equation 
and performing a Fourier transform we obtain an  equation for the phase $\phi_n(k;t)$, namely,
\begin{equation}\label{phase}
\dot \phi_n(q;t)+E(q)+E_n=-\int\frac{dk}{2\pi}eU(q-k,t){\rm e}^{i\phi_n(k;t)-i\phi_n(q;t)}.
\end{equation}
Here the dot indicates a derivative with regard to time, $E(q)=\hbar^2q^2/2m$
and $eU(q,t)$ is the Fourier transform (with regard to $x$) of $eU(x,t)$.
This equation has now to be integrated with the initial condition $\phi_n(k;t=0)=0$.
To do this we assume that  the potential $eU(k;t)$ is only a small perturbation 
so we can solve the equation perturbatively. To zero-th order in 
the potential the phase simply is $\phi_{n,0}(q,t)=-\left[E_n+E(q)\right]t$. In addition, for a right-moving
wavepacket only wave vectors close to the (positive) Fermi wave vector
$k_{F,n}=+\sqrt{2m(E_F-E_n)}/\hbar$ of channel $n$ are important.
Substituting $\phi_{n,0}(q,t)$ in the right hand side of Eq.~(\ref{phase}) 
and linearizing the phase around $k_{F,n}$ we get 
 $\phi_n(q;t)=\phi_{n,0}(k;t)+\delta\phi_n(t,0)$ where
\begin{equation}\label{deltaphi}
\delta\phi_{n}(t,t')=-\int_{t'}^t dt_1eU\left(v_{F,n}(t_1-t'),t_1\right)
\end{equation}
is independent of $q$. Finally, we can substitute the result for the phase
 into Eq.~(\ref{wavepacket}) and can then perform the $k$-integration
with the result
\begin{equation}
\Psi_n(x,t)=\delta(x-v_{F,n}t){\rm e}^{ik_{F,n}x-iE_Ft/\hbar-i\delta\phi_{n}(t,0)/\hbar}.
\end{equation}
Note that the particle remains well localized due to the linearization
around the Fermi point. 

From this result we can now read off  the scattering matrix $S(t,t')$ which connects
outgoing current amplitudes at point $x=L$ and time $t'$ to current amplitudes incoming
 at $x=0$ at time $t$.  With the proper normalization for the scattering matrix we obtain 
\begin{equation}\label{sarm}
(S_i)_{nn}(t,t')=\delta(t-t'-L_i/v_{F,n}){\rm e}^{ik_FL_i+i\delta \phi_{i,n}(t,t')/\hbar}.
\end{equation}
Here, $(S_i)_{nn}(t,t')$ are the diagonal elements of the scattering matrix.
 For later use we have introduced the additional index $i=A,B$ which we
 use to distinguishes  between the two arms of the ring. The phase $\delta \phi_{n,i}(t,t')$
 is found from Eq.~(\ref{deltaphi}) if we substitute $eU_i(x,t)$ for $eU(x,t)$.

\subsection{Average Conductance}
We are now in the position to  calculate the conductance matrix of the mesoscopic Mach-Zehnder 
interferometer. Quite generally, the statistically averaged linear response conductance of
a multi-terminal conductor is 
\begin{equation}\label{conductance}
\frac{G_{\alpha\beta}}{G_0}=-\left\langle\int_{-\infty}^\infty dt_1 dt_2 {\rm tr}\left[P_\alpha S(t,t_1) P_\beta S^\dagger(t_2,t)\right]\tilde f(t_1-t_2)\right\rangle_\phi.
\end{equation}
This is the multi-terminal equivalent of the expression given in \cite{vavilov,polianski}
for the conductance of a quantum dot. The angular brackets $\langle\ldots\rangle_\phi$ 
indicate statistical averaging over the fluctuations of the internal potentials
which renders the expression in Eq.~(\ref{conductance}) independent of time.
We assumed that a small voltage is applied to contact $\beta$ and $\alpha\neq\beta$.
 The total scattering matrix has the dimension
$M\times M$ where $M=\sum_\alpha N_\alpha$ and $N_\alpha$ is the number of channels
in contact $\alpha$. Here $G_0=2e^2/h$ is the conductance
 quantum and $P_\alpha$ is a diagonal matrix of dimension $M\times M$ projecting on 
contact $\alpha$ ($(P_1)_{ii}=1$, $1\leq i\leq N_1$ and $(P_1)_{ij}=0$ else, etc.). 
Furthermore, 
\begin{equation}
\tilde f(t)=\int_{-\infty}^\infty dE {\rm e}^{-iEt/\hbar}\frac{df(E)}{dE}=
\frac{\pi t/\hbar\beta}{\sinh\left(\pi t/\hbar\beta\right)}.
\end{equation}
is the Fourier transform of the derivative of the Fermi function $f(E)$.

 From now on we will concentrate on the single channel limit and will therefore
drop the channel index $n$. First, we consider the non-interacting case where there are no fluctuating 
potentials in the arms of the ring. From Eq.~(\ref{sarm}) we see that the scattering matrix 
then simply is a delta-function multiplied by a constant phase. If we now substitute
Eq.~(\ref{sarm}) into Eq.~(\ref{s31}) for the scattering matrix element $S_{31}(t,t')$ and
subsequently use Eq.~(\ref{conductance}) we get
\begin{equation}\label{thermal}
\frac{G_{31}}{G_0}=-2{\cal T}(1-{\cal T})\left[1+\left(\frac{kT}{E_\Delta}\right)\frac{\cos\left(k_F\Delta L+ 2\pi\Phi\right)}{\sinh\left(kT/E_\Delta\right)}\right].
\end{equation}
Here, $E_\Delta=\hbar v_F/(\pi\Delta L)$, $\Delta L=L_A-L_B$ and $T$ is the
temperature.  The result Eq.~(\ref{thermal}) is of course exactly the same
that we would also have obtained from time-independent scattering theory.

From Eq.~(\ref{thermal}) it is seen that the 
visibility of the AB oscillations decays exponentially  for temperatures $kT\gg E_\Delta$.
For an interferometer with two arms of exactly the  same length Eq.~(\ref{thermal})
 thus predicts that temperature has no effect.
Contrary to our simple model,  temperature averaging will be important in
an experiment.  In Ref.~\cite{hansen}  it is indicated how the
 effect of thermal smearing can be taken into account when one is interested
 in extracting  the decoherence rate from the experimentally accessible total
 decay rate due to decoherence and thermal smearing.  

From now on we will concentrate on the case $\Delta L=0$ of an interferometer 
with two arms of equal length where there is no thermal averaging
but we will include the effects of interaction-induced
 potential fluctuations in the arms or the interferometer.
Again, we use Eq.~(\ref{sarm}) (now including a random time-dependent phase)
 and Eq.~(\ref{s31}) to get the matrix element $S_{31}(t,t')$
for the interferometer. If we substitute into Eq.~(\ref{conductance}) for $G_{31}$
and then average the conductance over the random potential fluctuations we get
\begin{equation}\label{average}
\frac{G_{31}}{G_0}=-2{\cal T}(1-{\cal T})\left[1-{\rm e}^{-\frac{1}{2\hbar^2}\langle \delta\tilde\phi(t)^2\rangle_\phi}\cos(2\pi\Phi)\right].
\end{equation} 
Here $\delta\tilde\phi(t)$ is the difference between the fluctuating phases
picked up along the two different interfering pathes. Note again, that
upon averaging statistically the conductance will not depend on time anymore.  
In our derivation of Eq.~(\ref{average}) we have treated $\delta\tilde\phi(t)$
as a Gaussian random variable with vanishing mean. In terms of the 
difference $e\Delta U(x,t)=eU_A(x,t)-eU_B(x,t)$ between the potentials in the two arms
the phase $\delta\tilde\phi(t)$ is
\begin{equation}\label{phasedifference}
\delta\tilde\phi(t)=\delta\phi_A(t,t-\tau)-\delta\phi_B(t,t-\tau)=e\int_{-\tau}^0 dt_1\Delta U(v_F(\tau+t_1),t+t_1)
\end{equation}
where $\tau=L/v_F$ is the traversal time. 

Next, we want to express  the phase-correlator
$\langle \delta\tilde\phi(t)^2\rangle$ through the
spectral function of the potential fluctuations since it is this quantity
that can usually be calculated. The spectrum $\Sigma_{\Delta U\Delta U}(\omega;x_1,x_2)$ is
defined through the relation
\begin{equation}\label{spectrum}
2\pi\delta(\omega+\omega')\Sigma_{\Delta U\Delta U}(\omega;x_1,x_2)=\langle\Delta U(x_1,\omega)\Delta U(x_2,\omega')\rangle_\phi.
\end{equation}
Here $\Delta U(x,\omega)$ is the Fourier transform of $\Delta U(x,t)$ with regard to time.
With this definition and using Eq.~(\ref{phasedifference}) it is not difficult to see that 
\begin{equation}\label{spacedep}
\langle \delta\tilde\phi(t)^2\rangle_\phi=\frac{e^2}{v_F^2}\int_{-\infty}^{\infty}\frac{d\omega}{2\pi}
\int_0^Ldx_1dx_2\Sigma_{\Delta U\Delta U}(\omega;x_1,x_2){\rm e}^{-i\omega(x_1-x_2)/v_F}.
\end{equation}
To make further progress in the calculation of the average conductance Eq.~(\ref{average}) 
we thus need to know the spectral function  $\Sigma_{\Delta U\Delta U}(\omega;x_1,x_2)$
 for the potential fluctuations.

\subsection{Single Potential Approximation}\label{spa}
In this section we want to discuss a simple limit where the potential fluctuation
spectrum and thus the average conductance Eq.~\ref{average} can be calculated 
analytically. Following  our previous work \cite{seelig1,seelig2}, we assume that
 each arm of the ring can be described by a single space-independent internal
 potential $U_i(t)$. The arms of the ring are thus treated as effectively 
zero-dimensional objects.  The difference between the potentials in the two arms
 is $\Delta U(t)=U_A(t)-U_B(t)$  and the spectral function $\Sigma_{\Delta U\Delta U}(\omega)$ 
 becomes a function of a single argument only.
At the same time we also assume that the spectral function varies only slowly on the
scale set by the inverse traversal time $\tau^{-1}$. In this limit Eq.~(\ref{spacedep})
simplifies considerably and we obtain
\begin{equation}\label{rate}
\frac{1}{2\hbar^2}\langle \delta\tilde\phi(t)^2\rangle_\phi=\tau\Gamma_\phi,\,\,\,\,\Gamma_\phi=\frac{e^2}{2\hbar^2}\Sigma_{UU}(0)
\end{equation}
where $\Gamma_\phi$ is the decoherence rate.
The exponent entering Eq.~(\ref{average}) then is linear in the 
traversal time $\tau$ (see Eq.~(\ref{rate})). Moreover, with the approximations described
 above the decoherence rate is proportional to the zero-frequency limit of the
 potential fluctuation spectrum $\Sigma_{UU}(\omega)$. 

Next, we need to find the decoherence rate and thus
the zero-frequency spectral function $\Sigma_{UU}(0)$ for the MZI. 
 We do not calculate the spectral functions of the potential from time-dependent
 scattering theory  but it was shown that in the zero-frequency limit 
the fluctuation spectra can be calculated from the knowledge of the on-shell
scattering matrix $S(E;{U_i})$ and its derivatives with regard to the local potentials
$S(E;{U_i})/dU_i$ (see Ref.~\cite{seelig2} and references therein).
 Details of the calculation of the decoherence rate for a MZI can be found in 
Refs.~\cite{seelig1,seelig2}. Here, we will only  review the main results.

 Our investigation is motivated by a recent experiment by Kobayashi et al.~\cite{kobayashi}
 where the decoherence rate for a four-terminal mesoscopic interferometer
 was extracted from a measurement  of the four-terminal resistance 
$R_{\alpha\beta,\gamma\delta}=(V_\gamma-V_\delta)/I_\alpha$ 
($I_\beta=-I_\alpha$). The two contacts $\alpha, \beta$ are voltage
 biased and monitored by an  amp\`eremeter while the two contacts $\gamma, \delta$ 
are connected to a voltmeter.
A resistance measurement  
is termed {\it local} if the voltage probes are along the current path  
and is termed {\it non-local} if the voltage probes are far from the current path.  
For the conductor shown in Fig.~(1), $R_{14,23}$ is a local resistance, whereas  
$R_{12,34}$ is an example of a non-local resistance.  
We emphasize that the sample is the same, independent of the resistance  
measured: what changes is how the sample is connected to the current source  
and to the voltmeter.   The experiment of Ref.~\cite{kobayashi} gave two surprising results:
First, the amplitude of the Aharonov-Bohm oscillations was found to be at least 
an order of magnitude larger in the non-local geometry. Second, the rate of suppression 
of the coherent oscillations due to the combined effect of decoherence and thermal
 averaging was observed to be around three times larger in the local geometry.
This discrepancy is almost entirely due to the difference between the decoherence 
rates in the local and non-local setups.  

In Ref.~\cite{seelig2} we have calculated the decoherence
rates due to charge fluctuations for the local and non-local setups for a single channel MZI 
within the single potential approximation. We can distinguish between two sources
of charge fluctuations and consequently potential fluctuations in the interferometer:
 there are thermal fluctuations of the
charge that are always present and independent of the external probe configuration
\cite{seelig1} but  in addition there are also charge fluctuations induced through the 
fluctuations of the voltages applied to the voltage probes. The second type of fluctuations
is strongly probe configuration dependent and is at the origin of the different 
decoherence rates for the local and non-local setups.  
If all four contacts of the interferometer are connected to a zero-impedance external
 circuit kept at constant voltage (no voltage fluctuations) the decoherence rate  is  
\begin{equation}\label{rate0} 
\gamma^0_\phi=\frac{2kTe^2}{\hbar^2}\left(\frac{C_\mu}{C}\right)^2R_q. 
\end{equation} 
Here, $T$ is the temperature, $R_q=h/(4e^2)$ is the charge 
relaxation resistance for an MZI in the one-channel limit and  the electrochemical capacitance 
$C^{-1}_\mu=C^{-1}+(e^2D)^{-1}$   is the series combination of the geometrical  
capacitance and the  density of states \cite{ap1}. For simplicity we have assumed
 that the capacitances coupling the two arms of the ring to the gates are the same for 
both arms ($C_A=C_B=C$, see Fig.~\ref{mzi_figure}).
As was found experimentally in Refs.~\cite{hansen} and \cite{kobayashi} the decoherence
rate is linear in temperature.

 The next step is to include the effect of the external circuit: In our calculations 
 we  assumed that current probes are part of a zero-impedance external circuit and exhibit current
but no voltage fluctuations while voltage probes are part of an external 
circuit with infinite impedance and thus exhibit voltage fluctuations but no
current fluctuations. 
\begin{figure}
\begin{center}
\includegraphics[scale=0.35]{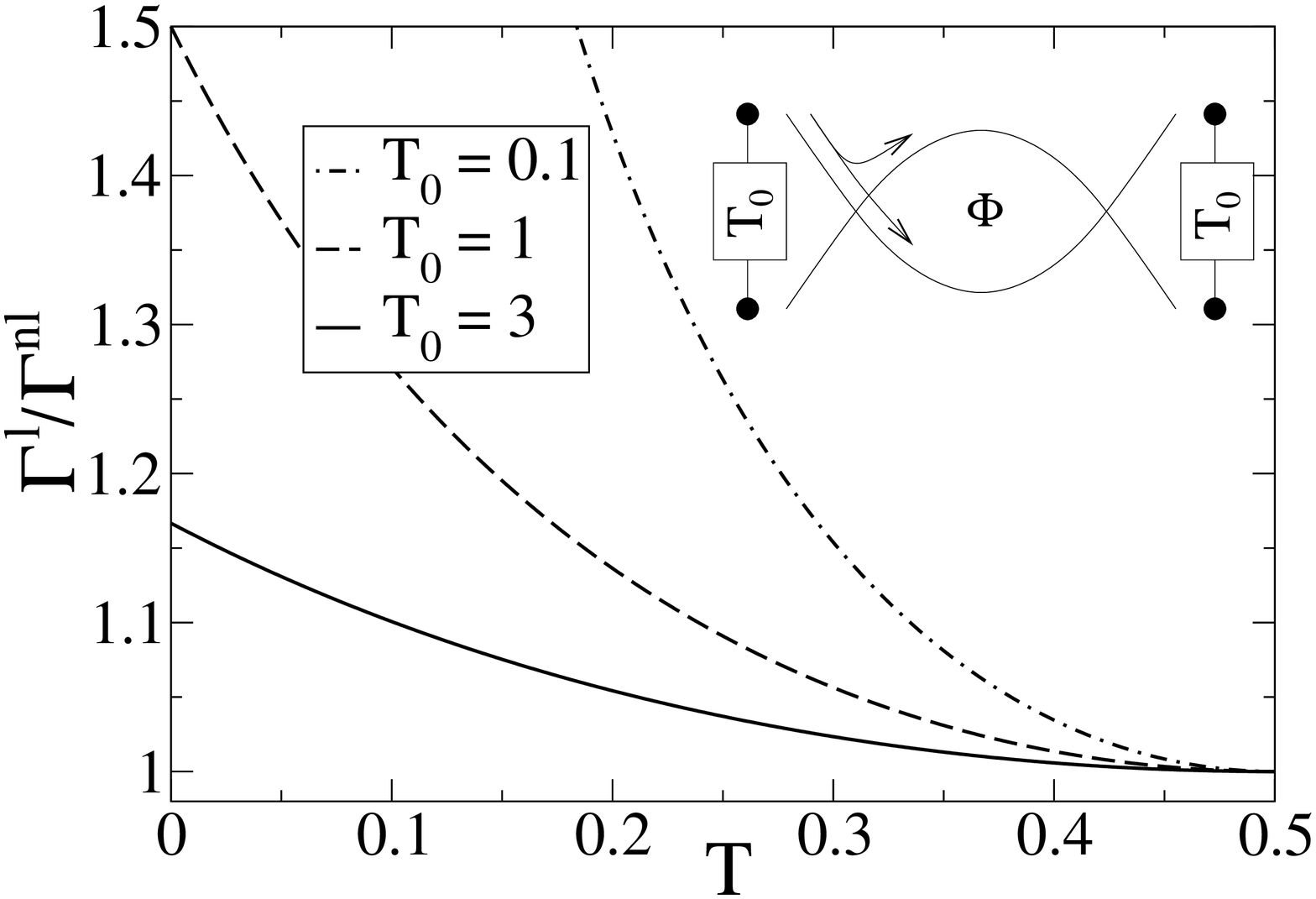}
\end{center}
\caption{The ratio of the local to the non-local decoherence rate 
is shown as a function of the  
transmission ${\cal T}$ at the beam splitters for different values of the incoherent
 parallel resistance $1/T_0$. All curves are symmetric with respect to ${\cal T}=1/2$. 
 In the inset, the two possible electron paths at the beam splitter 
 and the  resistance $1/T_0$ 
(in units of $h/e^2$) are indicated. After Ref.~\cite{seelig2}, copyright (2003) by the American Physical Society.} 
\label{rate_fig}
\end{figure}
With these assumptions we obtained for the dephasing rates 
in the local ($\it l$) and non-local ($\it nl$) configuration 
 respectively, 
\begin{subequations}
\label{gammas}
\begin{eqnarray} 
\Gamma^{l}_\phi=\gamma^0_\phi+\gamma^{l}_\phi,&&\,\,\,\gamma^{l}_\phi=\gamma^0_\phi\,\frac{(2{\cal T}-1)^2}{2{\cal T}(1-{\cal T})+T_0},\label{rate_local}\\ 
\Gamma^{nl}_\phi=\gamma^0_\phi+\gamma^{nl}_\phi,&&\,\,\,\gamma^{nl}_\phi=\gamma^0_\phi\,\frac{(2{\cal
    T}-1)^2}{1+2T_0}. \label{rate_nonlocal} 
\end{eqnarray}
\end{subequations}
Here, $\gamma^{l}_\phi$ and $\gamma^{nl}_\phi$ are  the probe-configuration 
specific contributions. 
The experiment of Ref.~\cite{kobayashi} shows 
 transmission between neighbouring contacts to be significant.
For better comparison, we therefore included a finite incoherent transmission
 $T_0=T_{12}=T_{21}=T_{34}=T_{43}$ (see inset Fig.~\ref{rate_fig}). 
In our calculations we have assumed that both beam splitters are identical
but relaxing this condition will not change the results qualitatively. 

The decoherence rate in the local (19a) and non-local configuration (19b) is
higher than in the voltage biased configuration (18). This can be understood
as an enhancement of the charge relaxation resistance $R_q$. Attaching a
high-ohmic voltmeter and thus closing some exits of the MZI will render charge
relaxation more difficult.

The decoherence rates Eq.~(\ref{gammas}) are strongly dependent on the
symmetry of the beam splitters. For perfectly symmetric beam splitters ${\cal
T}=1/2$ the two configurations are equivalent and the dephasing rate in both
cases is simply given by $\gamma^0_\phi$.  In the opposite limit of strong
asymmetry (${\cal T}\rightarrow 0$ or ${\cal T}\rightarrow 1$) however, the
local decoherence rate is greatly enhanced over the non-local decoherence rate
due to the symmetry-dependent denominator in the configuration specific part
of the dephasing rate. The divergence of the local decoherence rate for ${\cal
T}=0$ and ${\cal T}=1$ is cut off by an incoherent parallel transmission
$T_0$. The ratio of the local to the non-local decoherence rate is shown in
Fig.~\ref{rate_fig} for different values of $T_0$.  It should be emphasized
that the difference between the local and non-local configuration survives
even if the intersections are not treated as ideal beam splitters but exhibit
a certain degree of randomness \cite{seelig2}.

\section{An interferometer with chaotic scattering}

\begin{figure}
\begin{center}
\includegraphics[scale=0.7]{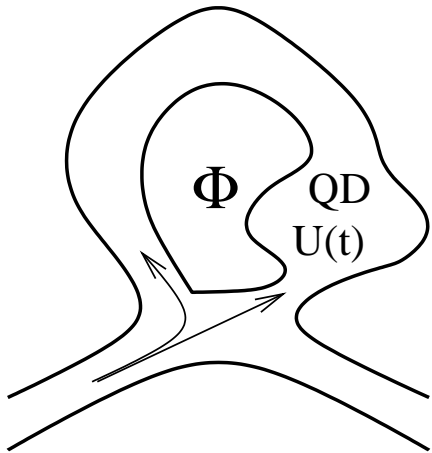}
\end{center}
\caption{Mesoscopic interferometer with a chaotic dot integrated in its arm.
The dot is capacitively coupled to a gate and interactions are taken into 
account through a self-consistent internal potential $U(t)$. The paths
an electron incoming from the left lead can take are also indicated.}
\label{ringdot}
\end{figure}

So far we have investigated the electronic equivalent of the
optical MZI and we have shown how charge fluctuations lead to decoherence.
In this section we want to consider a different type of interferometer 
shown in Fig.~\ref{ringdot}. As in the case of the MZI the region where the
 external leads are connected to the ring is  modelled by a beam splitter
 (see Eq.~(\ref{beamsplitter})). An electron coming from an  external lead 
will thus enter the ring with certainty and an electron arriving at
the intersection from the ring will go out into one of the reservoirs. This
implies that any given electron can go around the ring exactly once. 
An electron going through the ring clockwise picks up a phase $k_FL+2\pi \Phi$
 and an electron going through the ring counter-clockwise picks up a phase
  $k_FL-2\pi \Phi$ (for simplicity we have assumed this phase to be the same 
for all channels). Interference will  be due to electron-trajectories 
enclosing the  ring clockwise with trajectories going around the ring in 
the counter-clockwise direction. To mimic the backscattering and channel mixing 
that is always present in an experiment (see e.g. Ref.~\cite{hansen,kobayashi}) 
a chaotic dot is embedded into the arm of the ring.  
We will in the following assume that all interactions take place in the
quantum dot and no interactions take place in the ring.  
Restricting the interactions to the quantum dot can certainly be justified
if the dwell time in the dot is long compared to the traversal time for the ring.
If we furthermore assume that the hole in the ring is large 
compared to the surface of the chaotic dot, a magnetic flux of the order of one flux 
quantum through the ring will not break time-reversal invariance in the dot.
The (time-independent) scattering matrices for the dot are then distributed 
according to the circular orthogonal ensemble.

As in Sec.~\ref{spa} we will treat interaction effects on the level of the
single potential approximation. We thus characterize the quantum dot by an
internal potential $U(t)$ that has to be determined self-consistently.  For an
open chaotic dot in the many channel limit this approximation is well
justified \cite{slava,igor1}; it is in fact very common in the study of
Coulomb blockade problems. In the many-channel limit the potential can be
considered as a Gaussian random variable with a spectrum $\Sigma_{UU}(\omega)$
\cite{note}.  In the presence of a time-dependent internal potential we have
to use scattering matrices $S(t,t')$ with two time arguments to describe
scattering in the dot. Since we assume that no scattering takes place in the
ring, the scattering matrix for the ring is a diagonal matrix with elements
$S_{nn}(t,t')=\delta(t-t'-L/v_{F}){\rm e}^{ik_FL\pm 2i\pi \Phi}$ (see
Eq.~(\ref{sarm}), the positive (negative) sign is for an electron going around
the ring in the (counter-) clockwise direction and $n$ is the channel index).
The ensemble averaged conductance can then be calculated using the
time-dependent scattering matrix theory of Ref.~\cite{polianski}.  In this
calculation we will need correlators of the type $\langle
S_{ij}(t,t')S^\ast_{kl}(s,s')\rangle$ which can be found there. After
performing both an ensemble average and -- in addition -- a statistical
average over the (Gaussian) potential fluctuations we obtain for the
conductance
\begin{equation}\label{cond}
\frac{G}{G_0}=-\frac{N}{2}+\frac{(1-2{\cal T})^2}{4}
+A_\phi{\cal T}(1-{\cal T})\cos(4\pi\Phi).
\end{equation}
Here, we have assumed that an equal number of channels $N\gg 1$ is open in all
 the leads. From Eq.~(\ref{cond}) we see that on the ensemble average there is no coherent
 (flux-dependent) contribution to the leading order term of the conductance.
Furthermore, the fluctuating internal potential  only affects the flux-dependent
 part of the weak localization correction to the conductance.  
The amplitude 
\begin{equation}
A_\phi=\left\{ \begin{array}{ll}
                 \left(1+2\tau_d\Gamma_\phi \right)^{-1},& \tau \gg \tau_d,\\
                 \exp\left(-2\tau\Gamma_\phi\right),& \tau \ll \tau_d.
               \end{array} \right.
\end{equation}
depends on the strength of the decoherence and can in principle be obtained exactly.
Here, we concentrated on the  two interesting limits $\tau \gg \tau_d$ and $\tau \ll \tau_d$. 
where $\tau_d=h/(2N\Delta)$ is the dwell time, $\Delta$ is  the level spacing in the dot and
$\tau=L/v_F$ is the traversal time for the ring. The decoherence rate $\Gamma_\phi=e^2\Sigma_{UU}(0)/2\hbar^2$ is defined as in  Eq.~(\ref{rate}).

A single (time-dependent) potential $U(t)$ can not introduce a phase difference between
interfering trajectories  confined to a quantum dot and will therefore not suppress or
 dephase the weak localization correction to the conductance. 
To understand the experimental results for  dephasing in chaotic dots of 
Refs.~\cite{bird,clarke,huibers1} one thus has to go beyond the single
 potential approximation. In our more complicated model,
however, it is possible that electrons on different trajectories through the ring pass
through the dot at different times. In this situation, dephasing due to a single
fluctuating potential in the dot becomes effective. The time during which two interfering electrons
are exposed to different potentials $U(t)$ is given by the minimum of the dwell
time $\tau_d$ and the traversal time $\tau$. 
In the limit $\tau \gg \tau_d$ the amplitude $A_\phi$ becomes independent of the traversal time $\tau$
and therefore of the circumference $L$ of the ring. This is a consequence of our assumption 
that interactions are restricted to the dot. In the opposite limit, $\tau \ll \tau_d$,
the amplitude $A_\phi$ depends exponentially on the traversal
time but is independent of the dwell time.  

For zero magnetic field
and if $\tau\rightarrow 0$ and thus $A_\phi\rightarrow 1$
the system reduces to a simple chaotic dot  and the
 fluctuating potential has no effect whatsoever.  In this case,
the weak localization correction is $1/4$ as was first obtained from
a scattering matrix approach in Refs.~\cite{baranger1,jalabert}.
In the limits ${\cal T}\rightarrow 0$ or  ${\cal T}\rightarrow 1$ 
the system  corresponds to a simple quantum dot with two external leads 
even for arbitrary traversal time and magnetic field.
For strong dephasing, $A_\phi=0$, the weak localization correction 
of Eq.~(\ref{cond}) is $(1-2{\cal T})^2/4$. It can be obtained
from the incoherent addition of the four different paths
between the two external contacts. (Note, that even in this "incoherent" regime
coherent backscattering does take place in the dot). 

In an experiment it might be interesting to vary the location of the quantum dot.
Decoherence due to a single fluctuating potential should be strongly suppressed
in a perfectly symmetric interferometer where the dot is connected to the beamsplitter
 by two arms of exactly the same length. In such a setup time-reversal symmetry would 
not be broken and even the flux-dependent term would not be dephased.

The low-frequency dynamics of the charge fluctuations in the dot is governed by a charge 
relaxation time $\tau_{RC}=R_qC_\mu$. In a multi-channel chaotic dot the charge 
relaxation time is much shorter than the dwell time $\tau_d$
($\tau_{RC}\ll\tau_d$). For the interferometer considered here,  we have in addition  
$\tau_{RC}\ll\tau$. It is this separation of time scales that allows us to relate 
the decoherence rate to the zero-frequency limit of the spectral function only.
The potential fluctuation spectrum $\Sigma_{UU}(\omega)$ for a chaotic dot can be found 
with the help of Ref.~\cite{brouwer3} where the dynamic conductance matrix for a chaotic dot
 capacitively coupled  to a gate was calculated. The fluctuation dissipation theorem in 
the classical limit gives $\Sigma_{UU}(\omega)=2kT{\rm Re}G_{GG}(\omega)/(C^2\omega^2)$ 
where $G_{GG}(\omega)$ is the matrix element relating the current induced into the gate 
to a variation of the gate voltage.  In the zero-frequency limit
 the spectral function for the charge fluctuations in the dot is
 $\Sigma_{UU}(0)=2kTR_q(C_\mu/C)^2$. The charge relaxation resistance for a dot
 attached to two leads with $N$ channels each is $R_q=(h/e^2)/(2N)$. It is thus 
just the parallel addition of the resistances  $(h/e^2)/N$ of the individual contacts.
Furthermore, the electrochemical capacitance $C_\mu$ is the series addition of the
geometrical capacitance $C$ and the density of states $e^2D$ ($D=\Delta^{-1}$), namely
$C^{-1}_\mu=C^{-1}+(e^2D)^{-1}$ \cite{ap1}. In the experimentally relevant limit
$C_\mu/C\rightarrow 1$ we thus get
\begin{equation}
\Gamma_\phi=\frac{\pi kT}{\hbar N}.
\end{equation}
The dependence of the dephasing rate on the size of the contacts, i.e. on $N$,
reflects the fact that potential fluctuations in the dot are the smaller
the larger the contacts.

It is interesting to compare  Eq.~(\ref{cond}) with the result obtained if
dephasing is introduced via a  voltage probe model \cite{voltageprobe,baranger2,brouwer1}. 
In this model a fictitious voltage probe  is connected to the dot. It is imposed
that no net current flows at this terminal such that every outgoing electron 
has to be replaced by an incoming one. The two exchanged electrons have 
however no phase-relation which leads to a suppression of coherence in the dot.   
Using the diagrammatic theory of Ref.~\cite{brouwer2} it is possible to calculate
the average conductance of the system including the voltage probe
and we obtain
\begin{equation}
\frac{G}{G_0}=-\frac{N}{2}+\frac{2N}{2N+\gamma}\left[\frac{(1-2{\cal T})^2}{4}
+{\cal T}(1-{\cal T})\cos(4\pi\Phi)\right].
\end{equation}
Note that in the limits ${\cal T}\rightarrow 0$ and  ${\cal T}\rightarrow 1$
we correctly recover the result for the conductance of a simple chaotic dot.
In contrast to a single fluctuating potential the voltage probe leads
to decoherence of the full weak-localization correction.
In principle one could also attach the voltage probe directly to the ring
instead of attaching it to the quantum dot. In this case the result would 
be similar to Eq.~(\ref{cond}) since a voltage probe attached to the ring will
not suppress interference between trajectories in the dot.

\section{Conclusions}
In this work we have investigated charge-fluctuation-induced decoherence of
 AB oscillations in mesoscopic interferometers. 
 As a first example we looked at an interferometer without backscattering, 
the electronic equivalent of an optical Mach-Zehnder interferometer. 
Employing a time-dependent scattering approach we have been able to relate
the suppression of coherent Aharonov-Bohm oscillations to the spectrum 
of internal potential fluctuations. Then, specializing to the case where both arms
of the ring are described by a single potential each, we have explicitly 
calculated the decoherence rate. Within our model the decoherence rate is linear
in temperature as observed in Refs.~\cite{hansen,kobayashi}.
We also demonstrated that  the potential fluctuations and thus the decoherence rate can 
strongly depend on the external measurement circuit.
This provides an  explanation of the measurement configuration dependence of
 the decoherence rate in a four-terminal mesoscopic ring reported in Ref.~\cite{kobayashi}.
 
As a second example we studied a mesoscopic ring with a chaotic dot 
included in its arm. This has allowed us to take into account backscattering
and channel mixing. Also, contrary to the case of the MZI where we were mainly interested
in the single-channel limit, we have allowed for a large number of channels in 
the ring. Making the assumption that interactions take place mainly in the dot
and describing the dot by a single internal potential we could again relate 
the decoherence rate to the potential fluctuation spectrum. 

It is important to emphasize that the conductors considered here do not 
have to be charge neutral. To the contrary, an open  mesoscopic 
conductor can easily be charged up relative to its surroundings by exchanging
charge with the reservoirs. Charge exchange between conductor and reservoir
is an important source of dephasing. This is reflected in the dependence of the 
dephasing rates we have calculated above on the charge relaxation resistance $R_q$. 
The charge relaxation resistance is a measure of the energy dissipated as 
a charge carrier relaxes into a reservoir. Charge relaxation becomes easier
 the more channels are open in the leads connecting conductor and reservoirs.
In consequence deviations from charge neutrality are evened out quicker and
charge fluctuations become less pronounced. This, in turn leads to a smaller
 decoherence rate.  

There are many interesting venues leading beyond the material presented here.
Clearly, it would be interesting to go beyond  the single potential approximation 
 to determine the decoherence rate  for a  mesoscopic interferometer. 
 Such a generalization would become necessary  if one was interested to know how
 charge fluctuations suppress the weak localization correction to the
 conductance of a quantum dot. 

\section*{Acknowledgements} 
This work was supported by the Swiss National Science Foundation.

\end{document}